\documentclass{article}
\usepackage[T1]{fontenc} 
\usepackage[utf8]{inputenc} 
\usepackage{ismir,amsmath,cite,url}
\usepackage{graphicx}
\usepackage{color}
\usepackage{float}

\usepackage{booktabs}
\usepackage{tabularx}
\title{MoisesDB: A Dataset for Source Separation beyond 4-Stems}
\multauthor
{Igor Pereira \hspace{1cm} Felipe Araújo \hspace{1cm} Filip Korzeniowski \hspace{1cm} Richard Vogl}
{
Moises Systems Inc., Salt Lake City, USA.\\
{\tt\small igor@moises.ai}
}

\def\authorname{I. Pereira, F. Araújo, F. Korzeniowski, and R. Vogl}

\usepackage[bookmarks=false,pdfauthor={\authorname},pdfsubject={\papersubject},hidelinks]{hyperref}

\begin{document}

\maketitle

\begin{abstract}

In this paper, we introduce the MoisesDB dataset for musical source separation. It consists of 240 tracks from 45 artists, covering twelve musical genres. 
For each song, we provide its individual audio sources, organized in a two-level hierarchical taxonomy of stems. 
This will facilitate building and evaluating fine-grained source separation systems that go beyond the limitation of using four stems (drums, bass, other, and vocals) due to lack of data. 
To facilitate the adoption of this dataset, we publish an easy-to-use Python library to download, process and use MoisesDB. 
Alongside a thorough documentation and analysis of the dataset contents, this work provides baseline results for open-source separation models for varying separation granularities (four, five, and six stems), and discuss their results. 
\end{abstract}

\section{Introduction}\label{sec:introduction}
Source separation is the task of splitting an audio signal into separate signals for each signal source.
For music, the signal sources are the instruments that appear in the track, e.g.: guitar, bass, piano, drums, and vocals.

Music source separation is a relevant task within music information retrieval.
While it can be used as a pre-processing step for other tasks (e.g.\ voice separation for f0 tracking), source separation enables diverse applications on arbitrary music tracks that would need manual creation of stems otherwise.
For example, in the context of music education, the creation of play-along tracks for students, facilitating by-ear transcription of relevant instruments, or automatic creation of karaoke backing tracks.
Such applications are relevant for industry, as demonstrated by initiatives like the demixing challenges\footnote{\url{https://www.aicrowd.com/challenges/music-demixing-challenge-ismir-2021/sound-demixing-challenge-2023}}.

State-of-the-art source separation systems are usually built using neural-network-based machine learning systems, trained in a supervised way \cite{spleeter2020, Kong2021, Rouard2022}.
In order to train these systems, a large amount of training data is required.
For supervised approaches, the training data is represented by pairs of {\em i}.\ a mixed audio track and {\em ii}.\ a set of so-called stems that, when combined, recreate the audio track.
Stems are audio signals containing only one (or a group of related) sources, i.e.\ instruments.
A pair of one mixed track and its corresponding stems constitutes one training example.

Besides the large amount of manual work involved in any large-scale dataset creation, this kind of data is especially hard to come by for several reasons.
Whenever dealing with music audio data, legal issues may arise by collecting and sharing a dataset. 
The copy and distribution rights for most music are held by music publishers and record labels and are enforced rigorously. 
Obtaining the audio recordings for the individual instruments (stems) along with the final mix may expose recording, mixing, and mastering techniques of the recording studios, responsible for producing a track, which is why recording studios may oppose the publishing of stems in order to keep their trade secrets. 
Finally, processing, exporting, and organizing stems from recording projects (often from a digital audio workstation) is a considerable task. 
Usually, these recording projects are created without considering the requirement of exporting instrument stems.
All these factors hinder the creation and release of multitrack and stem datasets.

\begin{table} 
\footnotesize
\begin{tabular}{@{}lccc@{}}
\toprule
\textbf{Dataset} & \textbf{Year} & \textbf{No. of Tracks} & \textbf{Stems / Multitracks} \\ 
\midrule
MedleyDB \cite{medleydbv1}         & 2014          & 122                   & Multitracks                  \\
MedyleyDB-V2 \cite{medleydbv2}     & 2016          & 196                   & Multitracks                  \\
\midrule
DSD100 \cite{dsd100}          & 2015          & 100                   & 4 Stems                       \\
MUSDB18 \cite{musdb18}          & 2017          & 150                   & 4 Stems                       \\
MUSDB18-HQ \cite{musdb18-hq}      & 2019          & 150                   & 4 Stems                       \\
\midrule
MoisesDB          & 2023          & 240                   & Multitracks                  \\ 
\bottomrule
\end{tabular}
\caption{Overview of publicly released datasets for music source separation. The datasets are grouped according to the set of tracks they contain. For example, DSD100 is a subset of MUSDB18. Additionally, 46 songs from MedleyDB are also used in MUSDB18.} 
\label{table::datasets_comp}
\end{table}

\begin{figure}[t]
    \centering
    \includegraphics{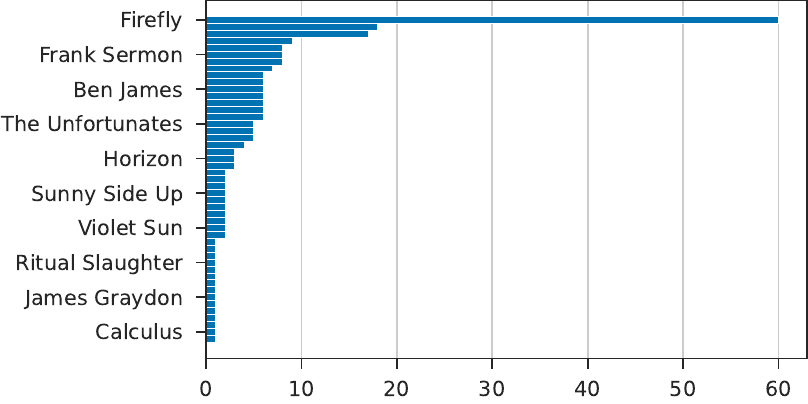}
    \caption{Artist distribution of MoisesDB.}
    \label{fig:artist_dist}
\end{figure}

\begin{figure}[h]
    \centering
    \includegraphics{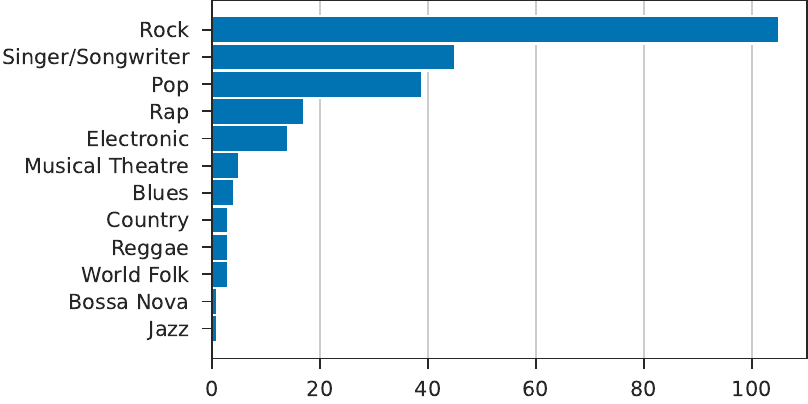}
    \caption{Genre distribution of MoisesDB.}
    \label{fig:genre_dist}
\end{figure}

While there exist source separation datasets aimed at a specific task, like vocal separation \cite{ikala2015, mir1k2010}, these are only of limited relevance for the more general task of splitting audio tracks up into stems. 
A majority of the existing stem datasets \cite{dsd100, musdb18-hq} use a limited taxonomy of four stems, namely: vocals, drums, bass, and other.
While this has become a de-facto standard for works on source separation \cite{spleeter2020, Kong2021, Rouard2022} due to the availability of data and comparability of results, this is a strong limitation of the resulting models.
For many practical applications, separation of other, widely used instruments may be relevant: e.g. guitars, keys, strings, etc.

Datasets featuring individually recorded tracks (multitrack, e.g. \cite{medleydbv2}), as well as other collections of multitrack recordings, like \emph{Open Multitrack Testbed} \cite{man2014open}, do exist.
However, these are not prepared to be used for source separation, out of the box, and may come with license restrictions.
Looking at recent source separation publications, we see that non-public data usually represents the bulk of training data (e.g. Bean dataset \cite{pretet2019singing} in  \cite{spleeter2020}; 800 tracks of undisclosed source in \cite{Rouard2022}).
This hints that by only using publicly available data, it is not possible to train competitive source separation models.
Thus, there is a need for more free data featuring a more detailed taxonomy, in order to be able to successfully train and test robust source separation models with the capability to separated more stems.

To improve the current situation, we introduce MoisesDB, a multitrack dataset featuring track annotations and a taxonomy to group individual tracks into stems. 
This dataset is offered free of charge for non-commercial research use only.
It consists of 240 music tracks from different artists and genres with a total duration of over 14 hours.
Along with the dataset, we provide baseline performance values for state-of-the-art source separation systems. 

The remainder of this work is structured as follows: 
Section~\ref{sec:prior_work} covers related work and contrasts it with the dataset presented here. 
Section~\ref{sec:dataset} discusses the details of MoisesDB.
Section~\ref{sec:performance} introduces baseline performance evaluation statistics using freely available source separation models.
Finally, Section~\ref{sec:conclusion} provides concluding remarks.

\begin{table}[t!]
\small
\centering
\begin{tabularx}{\columnwidth}{@{}lX@{}}
\toprule
\textbf{Stem} &                                                                               \textbf{Track} \\
\midrule
         Bass &                                                    Bass Guitar, Bass Synthesizer, Contrabass \vspace{0.4em} \\ 
Bowed Strings &               Cello, Cello Section, Other Strings, String Section, Viola Section, Viola Solo \vspace{0.4em} \\
        Drums & Cymbals, Drum Machine, Full Acoustic Drumkit, Hi-Hat, Kick Drum, Overheads, Snare Drum, Toms \vspace{0.4em} \\
       Guitar &                            Acoustic Guitar, Clean Electric Guitar, Distorted Electric Guitar \vspace{0.4em} \\
        Other &                                                                                           Fx \vspace{0.4em} \\
   Other Keys &                                   Organ, Electric Organ, Other Sounds, Synth Lead, Synth Pad \vspace{0.4em} \\
Other Plucked &                                                               Banjo/Mandolin/Ukulele/Harp   \vspace{0.4em} \\
   Percussion &                                                       A-Tonal Percussion, Pitched Percussion \vspace{0.4em} \\
        Piano &                                                                  Electric Piano, Grand Piano \vspace{0.4em} \\
       Vocals &                               Background Vocals, Lead Female Singer, Lead Male Singer, Other \vspace{0.4em} \\
         Wind &                                                             Brass, Flutes, Other Wind, Reeds \\
\bottomrule
\end{tabularx}
\caption{MoisesDB stem-track taxonomy used to organize individual tracks into stems.}
\label{tab:taxonomy}
\end{table}

\begin{figure}[t]
    \centering
    \includegraphics{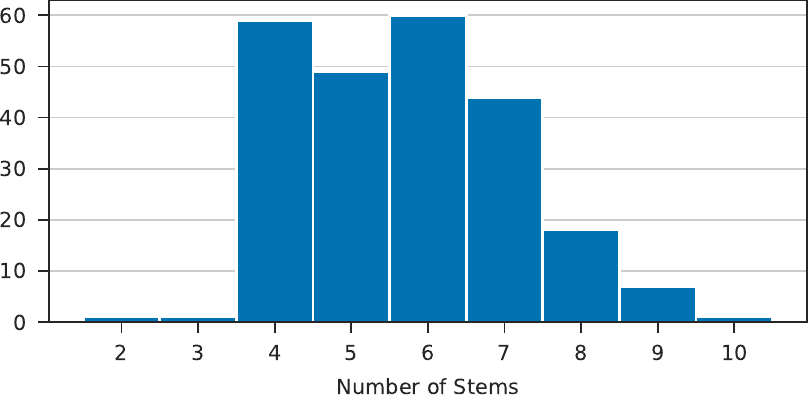}
    \caption{Number of stems per track in MoisesDB.}
    \label{fig:num_stem_dist}
\end{figure}

\section{Related Work}\label{sec:prior_work}

In the past, several multitrack and stem datasets have been published by the community (see Tab.~\ref{table::datasets_comp}).
This section will discuss their properties and set the context for the dataset presented in this work.
Since the main focus of this work is source separation into as many stems as possible, single stem focused datasets (e.g. voice separation datasets \cite{ikala2015, mir1k2010}) will be mainly ignored.

In 2014, Bittner et al.\ released the \emph{MedleyDB} dataset \cite{medleydbv1}, which comprises 122 songs in multitrack format. It was extended by 74 songs (totalling 196 songs) in 2016, and published as \emph{MedleyDB 2.0} \cite{medleydbv2}.
The dataset provides audio files in a hierarchical structure, where the final mix is split into multiple stems, each containing numerous raw audio sources (multitracks). 
Besides the multitrack data, the MedleyDB dataset provides an extensive list of metadata, such as artist, track name, origin, genre, and producer, amongst others. 
Additionally it provides multiple annotations, such as instrument activation, melody, and pitch.

\begin{figure}[t!]
    \centering
    \includegraphics{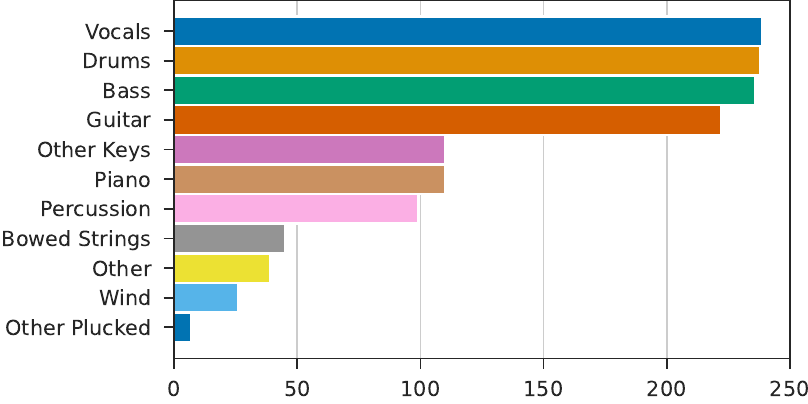}
    \caption{Distribution of stems in MoisesDB.}
    \label{fig:stem_dist}
\end{figure}

\begin{figure}[t]
    \begin{small}
    \begin{verbatim}
from moisesdb.dataset import MoisesDB

db = MoisesDB(data_path='./moises-db-data')
n_songs = len(db)
track = db[0]
# mix multitracks to stems
stems = track.stems  
# stems = {
#   'vocals': np.ndarray (stem audio data), 
#   'bass': np.ndarray (stem audio data), 
#   ...}
# mixture: np.ndarray
mixture = track.audio
# save mixed stems
track.save_stems('./stems/track_0')
    \end{verbatim}
    \end{small}
    \caption{Usage of the \texttt{MoisesDB} Python package.}
    \label{lst:moisesdb}
\end{figure}

\begin{figure}[h]
    \centering
    \includegraphics{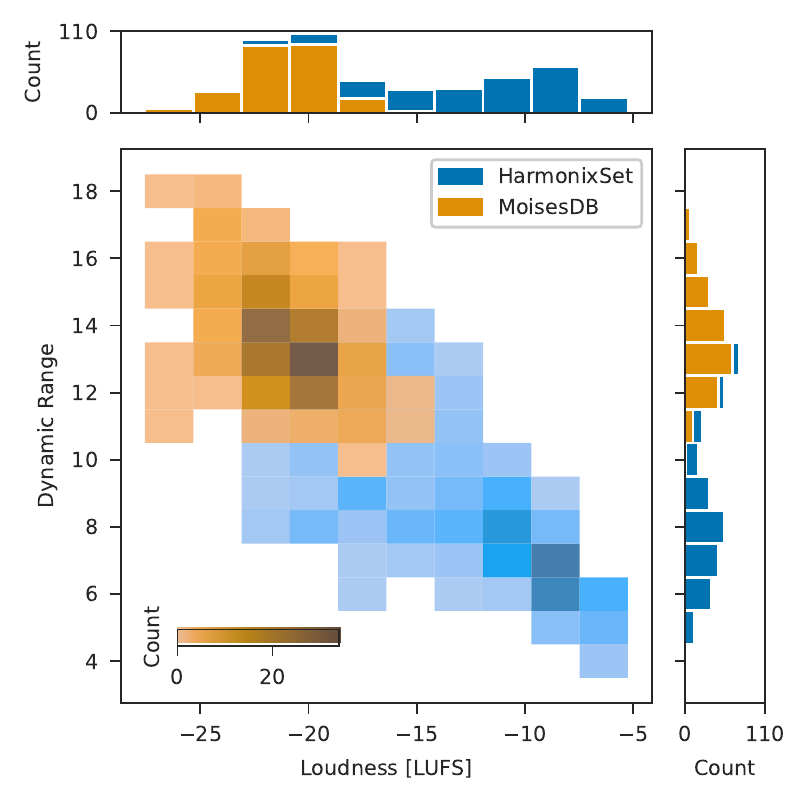}
    \caption{Loudness and Dynamic Range distribution of tracks in MoisesDB. For a comparison with commercially mixed and mastered songs, we sampled 240 tracks from the HarmonixSet~\cite{nieto_harmonix_2019}.
    }
    \label{fig:loudness_dr_dist}
\end{figure}

The annotations in MedleyDB make it useful for many MIR tasks, including the source separation of diverse instruments.
However, the shortcoming of MedleyDB for music source separation is the way it organizes tracks into stems. 
While it provides instrument information for each of them, and functional annotations for stems (such as ``melody'' or ``bass''), stems are not meaningfully labelled, only numbered. As a result, stem 01 of one song may be the drum kit, while stem 01 of a another mix is the bassoon. 
Furthermore, instruments---and thus tracks---are grouped according to how they physically produce their sound, rather than their role in the mix of a song.
For example, the ``drum machine'' falls into the same category as ``electric piano'', namely ``electric$\rightarrow$electronic''.
These shortcomings make it cumbersome to use for music source separation out of the box and significant work has to be done in order to use it for this task.

In 2016, Liutkus et al.\ released the \emph{DSD100} \cite{dsd100} dataset as part of the 2016 signal separation evaluation campaign to develop and benchmark source separation models. 
It contains 100 songs and uses the the four-stems taxonomy (vocals, drums, bass, and other).
Later, in 2017, Rafii et al. extended DSD100 to 150 songs by adding 46 pieces from MedleyDB, and including four previously unreleased recordings from commercial providers. This dataset became known as the \emph{MUSDB18} \cite{musdb18} dataset, and was used for the the 2018 signal separation evaluation campaign. 
In 2019, Z. Rafii et al. released an uncompressed version of the MUSDB18 dataset, MUSDB18-HQ \cite{musdb18-hq}.
As its predecessor DSD100, this dataset provides four stems---vocals, drums, bass, and other---as well as linear mixes.
MUSDB18 is widely used to train and benchmark source separation models, but the limited number of stems prevents researchers from building more granular source separation systems.  

In summary, data for training granular source separation systems is scarce: the 150 tracks from MUSDB18 are ready to use, but offer only four stems to separate; the 140 remaining tracks from MedleyDB (46 of the originally 196 are already part of MUSDB18) are not organized in a way that easily supports source separation research. 
This issue is also reflected in the fact that state-of-the-art source separation models often use larger, non-public datasets for training \cite{spleeter2020, Rouard2022}, or have to resort to synthetic training data (e.g.\cite{ozer2022source, jeon2022slakh}).
Other works find that MUSDB18's \emph{"source groupings remain overly coarse for many real-world remixing applications."} \cite{manilow2020hierarchical}.
To address these issues and to foster more research in music source separation, we created the MoisesDB dataset. 

MoisesDB comprises the largest publicly available set of multitrack audio recordings---240 previously unreleased songs---organized in a taxonomy that reflects the needs of source separation systems (as detailed in Sec.~\ref{sec:taxonomny}). The large number of songs, the diverse types of stems and tracks, and their organization in a source-separation-focused taxonomy will allow researchers to build their own stems according to their own requirements, and thus develop more granular source separation systems.

\begin{figure*}[t!]
    \centering
    \includegraphics[width=0.99\textwidth]{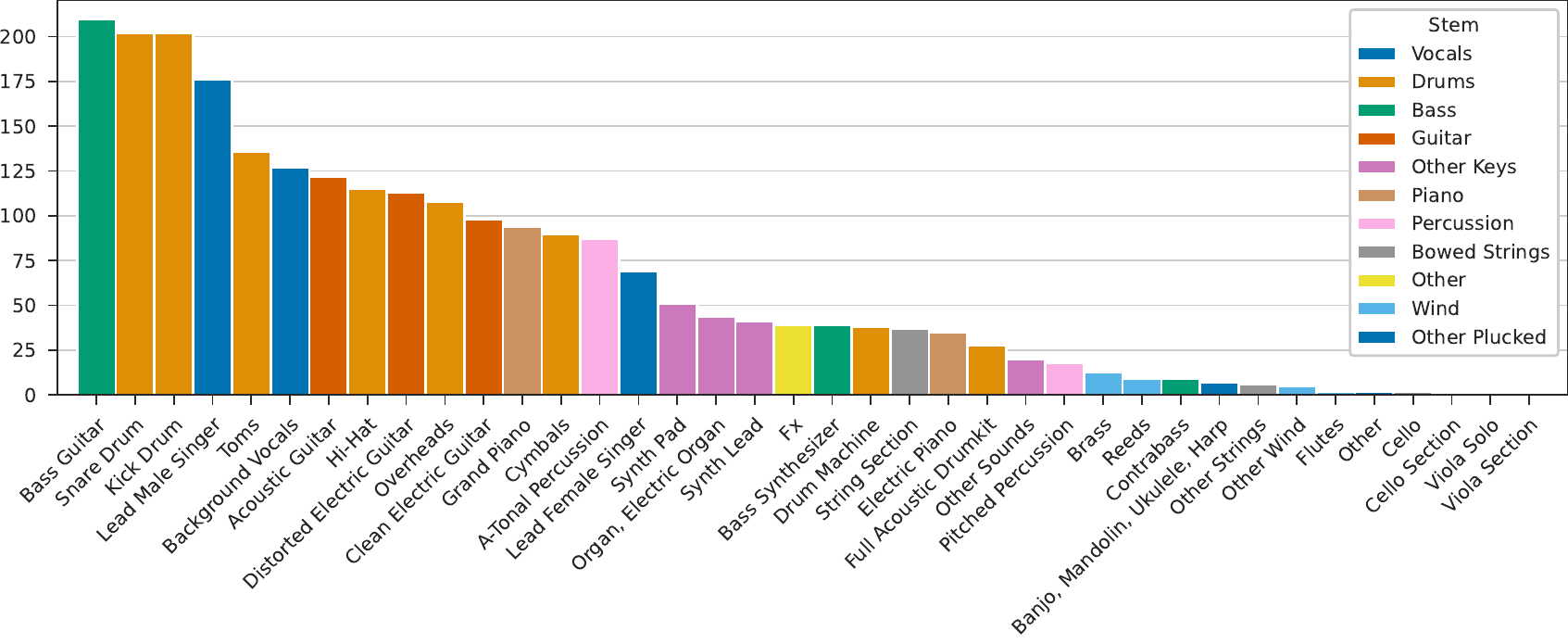}
    \caption{Distribution of tracks in MoisesDB.}
    \label{fig:sources_dist}
\end{figure*}

\section{Dataset}\label{sec:dataset}

MoisesDB consists of 240 songs by 47 artists that span twelve high-level genres. Both artists and genres follow a power-law-like distribution, where the majority of songs belong to few genres and are performed by few artists---see Fig.~\ref{fig:artist_dist} and \ref{fig:genre_dist}. The total duration of the dataset is 14 hours, 24 minutes and 46 seconds, where the average recording is 3:36 seconds, with a standard deviation of 66 seconds. 

\subsection{Stem Taxonomy}
\label{sec:taxonomny}

Modern song recordings consist of multiple recorded \emph{tracks}, which can be grouped and down-mixed into a smaller number of \emph{stems}. For example, the ``drums'' stem might comprise tracks for the snare drum, the bass drum, hi-hat, cymbals, and so on. MoisesDB provides all individual tracks for each song, grouped into stems by the taxonomy shown in Table~\ref{tab:taxonomy}. This taxonomy reflects the recording \& mixing process, and thus facilitates its reversal---music source separation---by grouping the raw tracks into semantically labeled stems. This also means that songs may consist of different numbers of stems, as shown in Fig.~\ref{fig:num_stem_dist}.
MoisesDB thus facilitates many future research directions: source separation models for a larger number of stems, data augmentation through mixing stems on-the-fly from their tracks, or separation of individual tracks from a stem, to name a few.

Given the genres of the songs in MoisesDB, certain stems are more common in the dataset than others: ``vocals'', ``drums'', and ``bass'' appear on virtually every song, while ``wind'' is rare. Similarly, certain tracks appear much more frequently than others, both within stems (``bass guitar'' vs. ``contrabass'') and between stems (``snare drum'' vs. ``cello''). Figs.~\ref{fig:stem_dist} and~\ref{fig:sources_dist} show the distributions of stems and sources, respectively. 

We anticipate that this imbalance will present a challenge in training source separation models for underrepresented stems, as it is likely that certain tracks, such as ``other plucked'' tracks, will still be difficult to distinguish from ``guitar'' tracks if trained solely on MoisesDB. 
However, the available data provides an opportunity for researchers to better identify and characterize errors made by their models. 
For instance, instead of simply observing that the separated ``other'' stem bleeds into ``guitar,'' MoisesDB enables researchers to pinpoint this issue to tracks where ``other'' includes plucked instruments.

\begin{figure*}
    \centering
    \includegraphics[width=\textwidth]{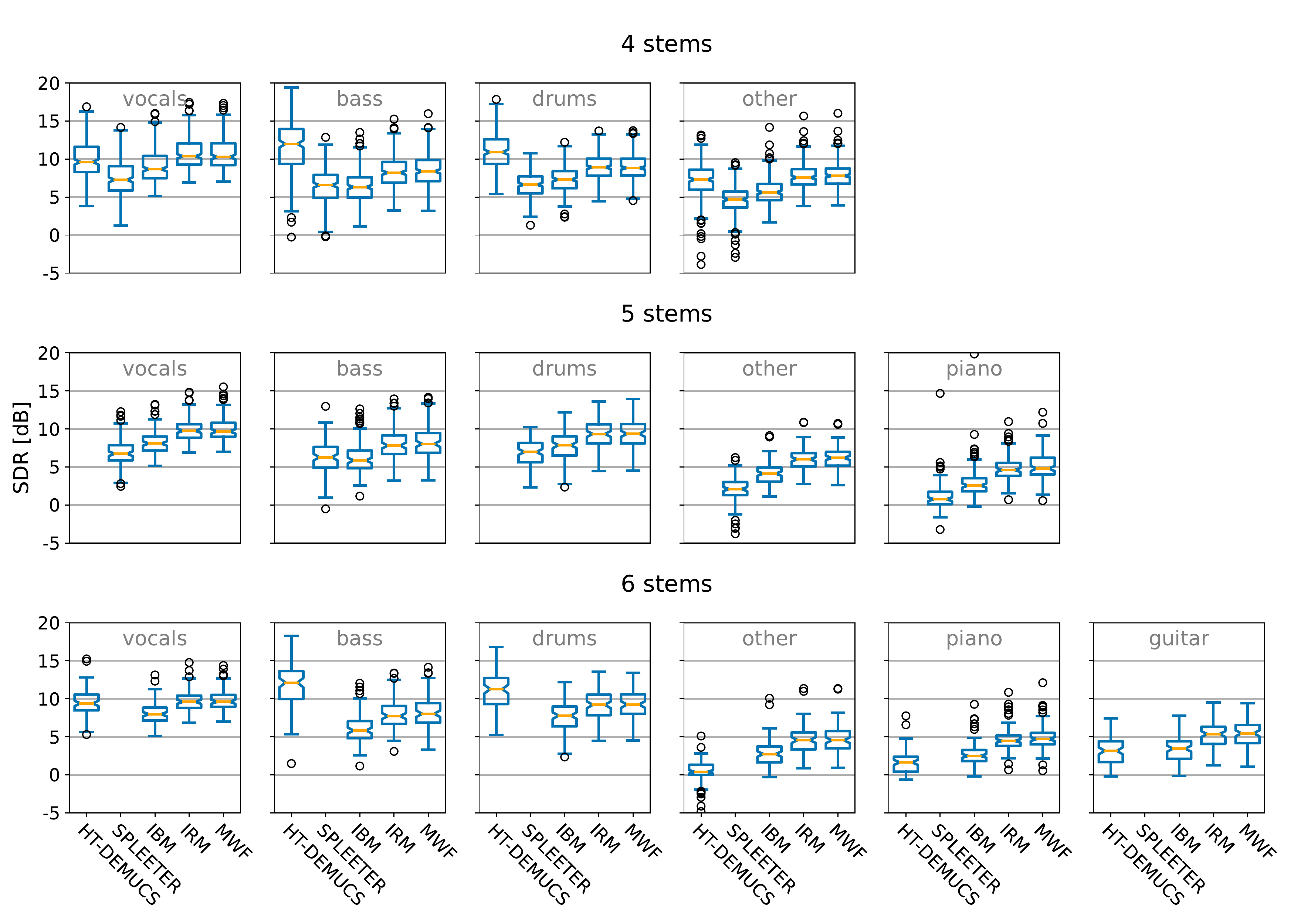}

    \caption{SDR values of each group of sources, for IBM, IRM, MWF, Demucs, and Spleeter source separation methods.}
    \label{fig:perf}
\end{figure*}
\begin{table*} 
\centering
\small
\begin{tabular}{@{}lrrrrrrrrrr@{}}
\toprule
\multicolumn{11}{c}{\textbf{4 stems (N = 235)}}                                                                                                                                                                                   \\ \midrule
\multicolumn{1}{r}{} & \multicolumn{2}{c}{\textbf{HT-Demucs}} & \multicolumn{2}{c}{\textbf{Spleeter}}  & \multicolumn{2}{c}{\textbf{IBM}}       & \multicolumn{2}{c}{\textbf{IRM}}       & \multicolumn{2}{c}{\textbf{MWF}}       \\
\multicolumn{1}{r}{} & \textbf{Mean $\pm$ Std} & \textbf{Mdn} & \textbf{Mean $\pm$ Std} & \textbf{Mdn} & \textbf{Mean $\pm$ Std} & \textbf{Mdn} & \textbf{Mean $\pm$ Std} & \textbf{Mdn} & \textbf{Mean $\pm$ Std} & \textbf{Mdn} \\ \midrule
vocals               & 10.05 $\pm$ 2.48        & 9.62         & 7.61 $\pm$ 2.45         & 7.27         & 9.02 $\pm$ 2.13         & 8.67         & \textbf{10.72 $\pm$ 2.03}        & 10.37        & \textbf{10.72 $\pm$ 2.11}        & 10.27        \\
bass                 & \textbf{11.64 $\pm$ 3.35}        & 11.99        & 6.46 $\pm$ 2.26         & 6.57         & 6.46 $\pm$ 2.08         & 6.31         & 8.43 $\pm$ 2.03         & 8.20         & 8.68 $\pm$ 2.07         & 8.38         \\
drums                & \textbf{10.94 $\pm$ 2.30}        & 10.91        & 6.65 $\pm$ 1.72         & 6.64         & 7.33 $\pm$ 1.77         & 7.30         & 8.98 $\pm$ 1.68         & 8.92         & 9.01 $\pm$ 1.67         & 8.83         \\
other                & 7.00 $\pm$ 2.76         & 7.30         & 4.45 $\pm$ 2.26         & 4.69         & 5.77 $\pm$ 1.72         & 5.61         & 7.74 $\pm$ 1.65         & 7.57         & \textbf{7.90 $\pm$ 1.65}         & 7.79         \\
\emph{overall}                  & \textbf{9.91 $\pm$ 3.27}         & 9.69         & 6.29 $\pm$ 2.47         & 6.24         & 7.14 $\pm$ 2.28         & 6.99         & 8.97 $\pm$ 2.16         & 8.81         & 9.08 $\pm$ 2.15         & 8.87         \\ \midrule
\multicolumn{11}{c}{\textbf{5 stems (N = 104)}}                                                                                                                                                                                   \\ \midrule
vocals               &                         &              & 6.99 $\pm$ 1.97         & 6.74         & 8.29 $\pm$ 1.66         & 8.08         & 9.94 $\pm$ 1.59         & 9.75         & \textbf{10.01 $\pm$ 1.71}        & 9.68         \\
bass                 &                         &              & 6.26 $\pm$ 2.27         & 6.28         & 6.13 $\pm$ 2.15         & 5.86         & 8.02 $\pm$ 2.07         & 7.82         & \textbf{8.32 $\pm$ 2.08}         & 8.03         \\
drums                &                         &              & 6.89 $\pm$ 1.88         & 6.97         & 7.67 $\pm$ 1.94         & 7.87         & 9.29 $\pm$ 1.84         & 9.34         & \textbf{9.32 $\pm$ 1.84}         & 9.36         \\
other                &                         &              & 1.97 $\pm$ 1.76         & 2.09         & 4.04 $\pm$ 1.47         & 4.13         & 6.00 $\pm$ 1.44         & 6.01         & \textbf{6.10 $\pm$ 1.48}         & 6.19         \\
piano                &                         &              & 1.17 $\pm$ 1.86         & 0.75         & 3.04 $\pm$ 2.37         & 2.55         & 4.99 $\pm$ 2.32         & 4.60         & \textbf{5.30 $\pm$ 2.46}         & 4.79         \\
\emph{overall}                  &                         &              & 4.66 $\pm$ 3.20         & 5.02         & 5.12 $\pm$ 2.81         & 4.87         & 7.65 $\pm$ 2.66         & 7.60         & \textbf{7.81 $\pm$ 2.66}         & 7.83         \\ \midrule
\multicolumn{11}{c}{\textbf{6 stems (N = 88)}}                                                                                                                                                                                    \\ \midrule
vocals               & 9.55 $\pm$ 1.87         & 9.39         &                         &              & 8.09 $\pm$ 1.51         & 7.98         & 9.73 $\pm$ 1.46         & 9.61         & \textbf{9.81 $\pm$ 1.49}         & 9.61         \\
bass                 & \textbf{11.93 $\pm$ 2.87}        & 12.13        &                         &              & 6.04 $\pm$ 1.98         & 5.83         & 7.92 $\pm$ 1.93         & 7.73         & 8.24 $\pm$ 1.96         & 8.03         \\
drums                & \textbf{11.02 $\pm$ 2.44}        & 11.28        &                         &              & 7.58 $\pm$ 1.96         & 7.79         & 9.19 $\pm$ 1.86         & 9.21         & 9.23 $\pm$ 1.85         & 9.25         \\
other                & 0.28 $\pm$ 1.84         & 0.39         &                         &              & 2.85 $\pm$ 1.76         & 2.74         & 4.67 $\pm$ 1.76         & 4.57         & \textbf{4.72 $\pm$ 1.82}         & 4.55         \\
piano                & 1.60 $\pm$ 1.68         & 1.64         &                         &              & 2.78 $\pm$ 1.61         & 2.49         & 4.71 $\pm$ 1.61         & 4.47         & \textbf{4.97 $\pm$ 1.74}         & 4.70         \\
guitar               & 3.07 $\pm$ 1.81         & 3.16         &                         &              & 3.35 $\pm$ 1.54         & 3.44         & 5.28 $\pm$ 1.54         & 5.36         & \textbf{5.41 $\pm$ 1.65}         & 5.46         \\
\emph{overall}                  & 6.24 $\pm$ 5.17         & 6.05         &                         &              & 5.12 $\pm$ 2.81         & 4.87         & 6.91 $\pm$ 2.70         & 6.69         & \textbf{7.06 $\pm$ 2.73}         & 6.89         \\ \bottomrule
\end{tabular}
\caption{Mean, standard deviation (Std), and median (Mdn) of the SDR in dB for each Model/Method and stem type. The varying number of available tracks is denoted by N. \emph{Overall} indicates performance over all tracks regardless of stem group. Best results are marked in bold.}
\label{tab:results}
\end{table*}

\subsection{Recording and Mastering}

The songs in MoisesDB are professionally recorded in stereo. The individual tracks are combined additively to create stems, which are then mixed together to produce the final version of the song. Due to technical limitations during recording, minuscule bleeding from other stems may be present for some of the tracks. No compression, equalization, or other effects are used during the mixing process, and the songs are not subjected to mastering. As a result, the song mixes have a lower loudness and a higher dynamic range than professionally mastered commercial songs.
This raises concerns about the distributional shift between un-mastered training data and commercial recordings.
Indeed, Jeon and Lee~\cite{jeon2022loud} have found that training separation models using mastered mixes can improve separation quality.
However, providing un-mastered mixes is common in existing datasets such as MUSDB18, and models such as HT-Demucs~\cite{Rouard2022} generalize reasonably well to mastered recordings, even if trained on un-mastered data.

Figure~\ref{fig:loudness_dr_dist} shows the loudness and dynamic range distributions for the dataset, where loudness is measured in LUFS (Loudness Units relative to Full Scale) \cite{itu-bs1770}, and Dynamic Range is computed based on the definitions of the ``Pleasurize Music Foundation'' as implemented in the ``DR14 T.meter'' software\footnote{\url{https://github.com/simon-r/dr14_t.meter}}.

\subsection{Python Library}

\noindent With MoisesDB comes a Python library that facilitates working with the dataset by parsing metadata and automatically building stems and mixes. Figure~\ref{lst:moisesdb} shows an example usage of the library. The code shown there initializes the library, retrieves the number of tracks, creates the stems and the full mix, and saves the individual stems to a directory. For a detailed and up-to-date documentation, we refer the reader to the GitHub repository\footnote{\label{fn:github}\url{https://github.com/moises-ai/moises-db}}.

\section{Benchmarking}\label{sec:performance}

In order to establish reference values for each track of the MoisesDB dataset, we computed the Source to Distortion Ratio (SDR) \cite{sdr_formula} scores for Ideal Binary Mask (IBM) \cite{ibm}, Ideal Ratio Mask (IRM) \cite{irm}, and Multichannel Wiener Filter (MWF) \cite{mwf} oracle separation methods. Additionally, we assessed SDR scores for two popular public available and open-source architectures: Hybrid Transformer Demucs (HT-DEMUCS) \cite{Rouard2022} and Spleeter \cite{spleeter2020}. 
The SDR scores were calculated for three different groups of sources: four, five, and six stems. Given the architecture of the open-source models, results for Spleeter are available for four and five stems, and for HT-DEMUCS for four and six stems. 

The SDR measure \cite{sdr_formula} represents how much of the energy in a true source signal is preserved in an estimated source signal after applying a separation algorithm. The equation can be defined as
\begin{align}
\mathrm{SDR} = 10 \log_{10}\frac{\sum_{n}\left|s(n)\right|^2+\epsilon}{\sum_{n}\left|s(n)-\hat{s}(n)\right|^2+\epsilon},
\end{align}
where $s(n)$ represents the true source signal at time $n$, $\hat{s}(n)$ represents the estimated source signal at time $n$, and the result is given in decibels (dB).

Table \ref{tab:results} shows the SDR values in dB for each group of stems (4, 5, and 6) evaluated in this benchmark. For a better comparison, we chose the stems available in the open-source models: vocals, bass, drums, other, guitar, and piano. We also pick tracks containing at least all the stems chosen for each group, which explains the distinct number of tracks in Table \ref{tab:results}. Songs with more individual tracks than the ones specified for each group were merged into the ``other'' stem using a linear sum strategy.

Figure \ref{fig:perf} depicts boxplots representing the distribution of the SDR metric for both oracle and separation methods, calculated for each group of tracks comprising 4, 5, and 6 stems. The groups of stems evaluated were vocal, bass, drums, other, piano, and guitar.
Detailed results for every track and each stem are provided in the GitHub\footref{fn:github} repository.

The first fact that calls our attention can be seen in Figure \ref{fig:perf}, where the SDR results of IRM and MWF oracle methods did not show a significant difference for all groups of stems. The striking fact is the performance of HT-DEMUCS architecture, which outperforms the oracle methods for bass and drums stems, for the groups of 4 and 6 stems tracks, as we can see in Figures \ref{fig:perf} A and C, respectively. Those results contrast with the slightly worse performance of HT-DEMUCS for other, piano, and guitar stems, compared with oracle methods, as seen in \figref{fig:perf}~C.

\newpage
\section{Conclusion}\label{sec:conclusion}

In this work, we introduced MoisesDB, a multitrack dataset with a hierarchical taxonomy aimed at more-than-four-stems source separation. We set the context by analysing the current landscape of source separation datasets and presented a comparison with other relevant datasets along with a detailed analysis of MoisesDB. 
Specifically, we discussed the organizational taxonomy focused on source separation, the distribution over track duration, the distribution over genres, and the number of songs for each stem and source available in the dataset.

Moreover, we include performance results for two publicly available source separation methods: HT-Demucs, which has the best overall SDR score evaluated on the MUSDB18 test set, and Spleeter, which was one of the first source separation models released and adopted by the general public. 
We also added results for a few masking-based oracle methods: IBM, IRM, and MWF, which indicate the theoretical performance limits for mask-based source separation models. 
Additionally, we provide an easy-to-use Python library to access the data which allows fast integration with machine learning libraries.

Overall, this paper represents a detailed report on the MoisesDB dataset, which will hopefully prove to be a great resource for the source separation community in the future. 
This work aims at facilitating the development of better and extended source separation models as well as providing opportunities to be applied for other use cases, such as automatic mixing and generative accompaniment systems, among others.

\clearpage
\bibliography{main}

\begin{thebibliography}{10}
\providecommand{\url}[1]{#1}
\csname url@samestyle\endcsname
\providecommand{\newblock}{\relax}
\providecommand{\bibinfo}[2]{#2}
\providecommand{\BIBentrySTDinterwordspacing}{\spaceskip=0pt\relax}
\providecommand{\BIBentryALTinterwordstretchfactor}{4}
\providecommand{\BIBentryALTinterwordspacing}{\spaceskip=\fontdimen2\font plus
\BIBentryALTinterwordstretchfactor\fontdimen3\font minus
  \fontdimen4\font\relax}
\providecommand{\BIBforeignlanguage}[2]{{%
\expandafter\ifx\csname l@#1\endcsname\relax
\typeout{** WARNING: IEEEtran.bst: No hyphenation pattern has been}%
\typeout{** loaded for the language `#1'. Using the pattern for}%
\typeout{** the default language instead.}%
\else
\language=\csname l@#1\endcsname
\fi
#2}}
\providecommand{\BIBdecl}{\relax}
\BIBdecl

\bibitem{spleeter2020}
R.~Hennequin, A.~Khlif, F.~Voituret, and M.~Moussallam, ``Spleeter: A fast and
  efficient music source separation tool with pre-trained models,''
  \emph{Journal of Open Source Software}, vol.~5, no.~50, p. 2154, Jun. 2020.

\bibitem{Kong2021}
Q.~Kong, Y.~Cao, H.~Liu, K.~Choi, and Y.~Wang, ``Decoupling magnitude and phase
  estimation with deep {ResUNet} for music source separation,''
  \emph{arXiv:2109.05418 [cs, eess]}, Sep. 2021.

\bibitem{Rouard2022}
S.~Rouard, F.~Massa, and A.~D{\'e}fossez, ``Hybrid transformers for music
  source separation,'' \emph{arXiv:2211.08553 [cs, eess]}, Nov. 2022.

\bibitem{medleydbv1}
R.~Bittner, J.~Salamon, M.~Tierney, M.~Mauch, C.~Cannam, and J.~Bello,
  ``{MedleyDB}: A multitrack dataset for annotation-intensive {MIR} research,''
  in \emph{Proceedings of the 15th International Society for Music Information
  Retrieval Conference (ISMIR)}, Taipei, Taiwan, Oct. 2014, pp. 155--160.

\bibitem{medleydbv2}
R.~Bittner, J.~Wilkins, H.~Yip, and J.~P. Bello, ``{MedleyDB} 2.0: New data and
  a system for sustainable data collection,'' in \emph{Late Breaking Demo of
  the 16th International Society for Music Information Retrieval Conference
  (ISMIR)}, New York, USA, Aug. 2016.

\bibitem{dsd100}
A.~Liutkus, F.-R. St{\"o}ter, Z.~Rafii, D.~Kitamura, B.~Rivet, N.~Ito, N.~Ono,
  and J.~Fontecave, ``The 2016 signal separation evaluation campaign,'' in
  \emph{Proceedings of the 13th International Conference on Latent Variable
  Analysis and Signal Separation (LVA/ICA)}, Grenoble, France, Feb. 2017, pp.
  323--332.

\bibitem{musdb18}
\BIBentryALTinterwordspacing
Z.~Rafii, A.~Liutkus, F.-R. St{\"o}ter, S.~I. Mimilakis, and R.~Bittner, ``The
  {MUSDB18} corpus for music separation,'' Dec. 2017. [Online]. Available:
  \url{https://doi.org/10.5281/zenodo.1117372}
\BIBentrySTDinterwordspacing

\bibitem{musdb18-hq}
\BIBentryALTinterwordspacing
------, ``{MUSDB18-HQ} - an uncompressed version of {MUSDB18},'' Dec. 2019.
  [Online]. Available: \url{https://doi.org/10.5281/zenodo.3338373}
\BIBentrySTDinterwordspacing

\bibitem{ikala2015}
T.-S. Chan, T.-C. Yeh, Z.-C. Fan, H.-W. Chen, L.~Su, Y.-H. Yang, and R.~Jang,
  ``Vocal activity informed singing voice separation with the ikala dataset,''
  in \emph{Proceedings of the 40th IEEE International Conference on Acoustics,
  Speech and Signal Processing (ICASSP)}, South Brisbane, QLD, Australia, Apr.
  2015, pp. 718--722.

\bibitem{mir1k2010}
C.-L. Hsu and J.-S.~R. Jang, ``On the improvement of singing voice separation
  for monaural recordings using the mir-1k dataset,'' \emph{IEEE Transactions
  on Audio, Speech, and Language Processing}, vol.~18, no.~2, pp. 310--319,
  2010.

\bibitem{man2014open}
B.~De~Man, M.~Mora-Mcginity, G.~Fazekas, and J.~D. Reiss, ``The open multitrack
  testbed,'' in \emph{Proceedings of the 137th Audio Engineering Society
  Convention}.\hskip 1em plus 0.5em minus 0.4em\relax Los Angeles, CA, USA:
  Audio Engineering Society, Oct. 2014.

\bibitem{pretet2019singing}
L.~Pr{\'e}tet, R.~Hennequin, J.~Royo-Letelier, and A.~Vaglio, ``Singing voice
  separation: A study on training data,'' in \emph{Proceedings of the IEEE
  International Conference on Acoustics, Speech and Signal Processing
  (ICASSP)}.\hskip 1em plus 0.5em minus 0.4em\relax Brighton, UK: IEEE, May
  2019, pp. 506--510.

\bibitem{nieto_harmonix_2019}
O.~Nieto, M.~{McCallum}, M.~E.~P. Davies, A.~Robertson, A.~Stark, and E.~Egozy,
  ``The {HARMONIX} set: Beats, downbeats, and functional segment annotations of
  western popular music,'' in \emph{Proceedings of the 20th International
  Society for Music Information Retrieval Conference ({ISMIR})}, pp. 565--572.

\bibitem{ozer2022source}
Y.~{\"O}zer and M.~M{\"u}ller, ``Source separation of piano concertos with
  test-time adaptation,'' in \emph{Proceedings of the 23rd International
  Society for Music Information Retrieval Conference (ISMIR)}, Bengaluru,
  India, Dec. 2022, pp. 493--500.

\bibitem{jeon2022slakh}
E.~Manilow, G.~Wichern, P.~Seetharaman, and J.~Le~Roux, ``Cutting music source
  separation some slakh: A dataset to study the impact of training data quality
  and quantity,'' in \emph{IEEE Workshop on Applications of Signal Processing
  to Audio and Acoustics (WASPAA)}.\hskip 1em plus 0.5em minus 0.4em\relax
  IEEE, Oct. 2019, pp. 45--49.

\bibitem{manilow2020hierarchical}
E.~Manilow, G.~Wichern, and J.~Le~Roux, ``Hierarchical musical instrument
  separation.'' in \emph{Proceedings of the 21st International Society for
  Music Information Retrieval Conference (ISMIR)}, Virtual Conference, Oct
  2020, pp. 376--383.

\bibitem{jeon2022loud}
C.-B. Jeon and K.~Lee, ``Towards robust music source separation on loud
  commercial music,'' in \emph{Proceedings of the 23rd International Society
  for Music Information Retrieval Conference (ISMIR)}, Bengaluru, India, Dec.
  2022, pp. 575--582.

\bibitem{itu-bs1770}
{ITU-R}, ``Algorithms to measure audio programme loudness and true-peak audio
  level,'' {International Telecommunication Union}, Recommendation BS.1770-4,
  Oct. 2015.

\bibitem{sdr_formula}
E.~Vincent, R.~Gribonval, and C.~Fevotte, ``Performance measurement in blind
  audio source separation,'' \emph{IEEE Transactions on Audio, Speech, and
  Language Processing}, vol.~14, no.~4, pp. 1462--1469, Jul. 2006.

\bibitem{ibm}
E.~Vincent, H.~Sawada, P.~Bofill, S.~Makino, and J.~P. Rosca, ``First stereo
  audio source separation evaluation campaign: Data, algorithms and results,''
  in \emph{Proceedings of the 7th International Conference on Independent
  Component Analysis and Signal Separation (ICA)}, London, UK, Sep. 2007, pp.
  552--559.

\bibitem{irm}
A.~Liutkus and R.~Badeau, ``Generalized {Wiener} filtering with fractional
  power spectrograms,'' in \emph{Proceedings of the IEEE International
  Conference on Acoustics, Speech and Signal Processing (ICASSP)}, South
  Brisbane, Australia, Apr. 2015, pp. 266--270.

\bibitem{mwf}
N.~Q.~K. Duong, E.~Vincent, and R.~Gribonval, ``Under-determined reverberant
  audio source separation using a full-rank spatial covariance model,''
  \emph{IEEE Transactions on Audio, Speech, and Language Processing}, vol.~18,
  no.~7, pp. 1830--1840, Sep. 2010.

\end{thebibliography}

\end{document}